\documentstyle[wstwocl,epsfig]{article}
\begin{document}
\bibliographystyle{unsrt}    
\newcommand{\st}{\scriptstyle}
\newcommand{\sst}{\scriptscriptstyle}
\newcommand{\mco}{\multicolumn}
\newcommand{\epp}{\epsilon^{\prime}}
\newcommand{\vep}{\varepsilon}
\newcommand{\ko}{K^0}
\newcommand{\kb}{\bar{K^0}}
\newcommand{\bo}{B^0}
\newcommand{\bb}{\bar{B^0}}
\newcommand{\al}{\alpha}
\newcommand{\ab}{\bar{\alpha}}
\newcommand{\tb}{{\tan \beta}}
\def\be{\begin{equation}}
\def\ee{\end{equation}}
\def\bea{\begin{eqnarray}}
\def\eea{\end{eqnarray}}
\def\atu{{\alpha_t^U}}
\def\cm{{\cal M}}
\def\dvc{{\Delta V_{cosm}}}
\def\ov{\overline}
\def\gev{{\rm \; GeV}}
\def\tev{{\rm \; TeV}}
\def\mpla{{M_{\rm P}}}
\def\msu{{M_{\rm SUSY}}}
\def\simlt{\stackrel{<}{{}_\sim}}
\def\cg{{\cal G}}
\def\str{{\rm \; Str \;}}
\def\mun{{M_{\rm U}}}
\def\zzbar{{(z, \overline{z})}}
\def\ap#1#2#3   {{\em Ann. Phys. (NY)} {\bf#1} (#2) #3}
\def\apj#1#2#3  {{\em Astrophys. J.} {\bf#1} (#2) #3}
\def\apjl#1#2#3 {{\em Astrophys. J. Lett.} {\bf#1} (#2) #3}
\def\app#1#2#3  {{\em Acta. Phys. Pol.} {\bf#1} (#2) #3}
\def\ar#1#2#3   {{\em Ann. Rev. Nucl. Part. Sci.} {\bf#1} (#2) #3}
\def\cpc#1#2#3  {{\em Computer Phys. Comm.} {\bf#1} (#2) #3}
\def\err#1#2#3  {{\it Erratum} {\bf#1} (#2) #3}
\def\ib#1#2#3   {{\it ibid.} {\bf#1} (#2) #3}
\def\jmp#1#2#3  {{\em J. Math. Phys.} {\bf#1} (#2) #3}
\def\ijmp#1#2#3 {{\em Int. J. Mod. Phys.} {\bf#1} (#2) #3}
\def\jetp#1#2#3 {{\em JETP Lett.} {\bf#1} (#2) #3}
\def\jpg#1#2#3  {{\em J. Phys. G.} {\bf#1} (#2) #3}
\def\mpl#1#2#3  {{\em Mod. Phys. Lett.} {\bf#1} (#2) #3}
\def\nat#1#2#3  {{\em Nature (London)} {\bf#1} (#2) #3}
\def\nc#1#2#3   {{\em Nuovo Cim.} {\bf#1} (#2) #3}
\def\nim#1#2#3  {{\em Nucl. Instr. Meth.} {\bf#1} (#2) #3}
\def\np#1#2#3   {{\em Nucl. Phys.} {\bf#1} (#2) #3}
\def\pcps#1#2#3 {{\em Proc. Cam. Phil. Soc.} {\bf#1} (#2) #3}
\def\pl#1#2#3   {{\em Phys. Lett.} {\bf#1} (#2) #3}
\def\prep#1#2#3 {{\em Phys. Rep.} {\bf#1} (#2) #3}
\def\prev#1#2#3 {{\em Phys. Rev.} {\bf#1} (#2) #3}
\def\prl#1#2#3  {{\em Phys. Rev. Lett.} {\bf#1} (#2) #3}
\def\prs#1#2#3  {{\em Proc. Roy. Soc.} {\bf#1} (#2) #3}
\def\ptp#1#2#3  {{\em Prog. Th. Phys.} {\bf#1} (#2) #3}
\def\ps#1#2#3   {{\em Physica Scripta} {\bf#1} (#2) #3}
\def\rmp#1#2#3  {{\em Rev. Mod. Phys.} {\bf#1} (#2) #3}
\def\rpp#1#2#3  {{\em Rep. Prog. Phys.} {\bf#1} (#2) #3}
\def\sjnp#1#2#3 {{\em Sov. J. Nucl. Phys.} {\bf#1} (#2) #3}
\def\spj#1#2#3  {{\em Sov. Phys. JEPT} {\bf#1} (#2) #3}
\def\spu#1#2#3  {{\em Sov. Phys.-Usp.} {\bf#1} (#2) #3}
\def\zp#1#2#3   {{\em Zeit. Phys.} {\bf#1} (#2) #3}
\setcounter{secnumdepth}{2} 
%
%
\title{EXTENSIONS OF THE STANDARD MODEL}

\firstauthors{Fabio Zwirner}

\firstaddress{Theory Division, CERN, CH-1211 Geneva 23,
Switzerland}
\secondaddress{INFN, Sezione di Padova,
Via Marzolo~8, I-35131 Padua, Italy}

\twocolumn[\maketitle\abstracts{
This talk begins with a brief general introduction to the
extensions of the Standard Model, reviewing the ideology
of effective field theories and its practical implications.
The central part deals with candidate extensions near the
Fermi scale, focusing on some phenomenological aspects of
the Minimal Supersymmetric Standard Model. The final part
discusses some possible low-energy implications of further
extensions near the Planck scale, namely superstring theories.}]

\section{Preamble (some facts and some ideology)}

It is quite obvious that the Standard Model (SM) {\bf must} be
extended. Among the `hard' arguments supporting the previous
statement, the strongest one is the fact that the SM does not
include a quantum theory of gravitational interactions.
Immediately after, one can mention the fact that some of the SM
couplings are not asymptotically free, making it almost surely
inconsistent as a formal Quantum Field Theory. One can add to
the above the usual `soft' argument that the SM has about 20
arbitrary parameters, which may seem too many for a fundamental
theory.

Whilst this does not give us direct information on the form
of the required SM extensions, it brings along an important
conceptual implication: the SM should be seen as an {\bf
effective field theory}, valid up to some physical cut-off
scale $\Lambda$. The basic rule of the game\cite{effective}
is to write down the most general local Lagrangian compatible
with the SM symmetries [i.e. the $SU(3) \times SU(2) \times
U(1)$ gauge symmetry and the Poincar\'e symmetry], scaling all
dimensionful couplings by appropriate powers of $\Lambda$.
The resulting dimensionless coefficients are then to be
interpreted as parameters, which can be either fitted to
experimental data or (if one is able to do so) theoretically
determined from the fundamental theory replacing the SM at the
scale $\Lambda$. Very schematically (and omitting all coefficients
and indices, as well as many theoretical subtleties, such as the
problems in regularizing chiral gauge theories):
\begin{eqnarray}
{\cal L}_{eff} & = &
\Lambda^4 + \Lambda^2 \Phi^2 \nonumber \\
& + &
\left( D \Phi \right)^2 + \overline{\Psi} \not{\! \! D}
\Psi + F^2 + \overline{\Psi} \Psi \Phi + \Phi^4 \nonumber \\
& +  &
{\overline{\Psi} \Psi \Phi \Phi \over \Lambda}
+
{\overline{\Psi} \Psi \overline{\Psi} \Psi \over \Lambda^2}
+ \ldots \, ,
\label{leff}
\end{eqnarray}
where $\Psi$ stands for the generic quark or lepton field,
$\Phi$ for the SM Higgs field, and $F$ for the field strength
of the SM gauge fields. The first line of eq.~(\ref{leff})
contains two operators carrying positive powers of $\Lambda$,
a cosmological constant term proportional to $\Lambda^4$ and
a scalar mass term proportional to $\Lambda^2$. Barring for
the moment the discussion of the cosmological constant term,
which becomes relevant only when the model is coupled to gravity,
it is important to observe that {\bf no} quantum SM symmetry is
recovered by setting to zero the coefficient of the scalar mass
term. On the contrary, the SM gauge invariance forbids
fermion mass terms of the form $\Lambda \overline{\Psi}
\Psi$. The second line of eq.~(\ref{leff}) contains
operators with no power-like dependence on $\Lambda$,
but only a milder, logarithmic dependence, due to infrared
renormalization effects. The operators of dimension $d \le 4$
exhibit two remarkable properties: all those allowed by the
symmetries are actually present in the SM; both baryon number
and the individual lepton numbers are automatically conserved.
The third line of eq.~(\ref{leff}) is indeed the starting
point of an expansion in inverse powers of $\Lambda$,
containing infinitely many terms. For energies and field VEVs
much smaller than $\Lambda$, the effects of these operators are
suppressed, and the physically most interesting ones are those
that violate some accidental symmetries of the $d \le 4$
operators. For example, a $d=5$ operator of the form $\overline{\Psi}
\Psi \Phi \Phi$ can generate a lepton-number-violating
Majorana neutrino mass of order $G_F^{-1} / \Lambda$ (where
$G_F^{-1/2} \simeq 300$~GeV is the Fermi scale), as in the
see-saw mechanism; some of the $d=6$ four-fermion operators can be
associated with flavour-changing neutral currents (FCNC) or with
baryon- and lepton-number-violating processes such as proton decay.

At this point, the question that naturally emerges is the following:
where is the cut-off scale $\Lambda$, at which the expansion of
eq.~(\ref{leff})  loses validity and the SM must be replaced by
a more fundamental theory? Two extreme but plausible answers can
be given:
\begin{description}
\item[{\bf (I)}]
$\Lambda$ is not much below the Planck scale, $\mpla \equiv G_N^{-1/2}
/ \sqrt{8 \pi} \simeq 2.4 \times 10^{18}$~GeV, as roughly suggested
by the measured strength of the fundamental interactions, including
the gravitational ones.
\item[{\bf (II)}]
$\Lambda$ is not much above the Fermi scale, as suggested by the
idea that new physics must be associated with the mechanism of
electroweak symmetry breaking.
\end{description}
In the absence of an explicit realization at a fundamental level,
each of the above answers can be heavily criticized. The criticism
of (I) has to do with the existence of the `quadratically divergent'
scalar mass operator, which becomes more and more `unnatural' as
$\Lambda$ increases above the electroweak scale\cite{natural}.
On general theoretical grounds, we would expect for such operator
a coefficient of order 1, but experimentally we need a strongly
suppressed coefficient, of order $G_F^{-1} / \Lambda^2$. However,
after taking into account quantum corrections, this coefficient
can be conceptually decomposed into the sum of two separate
contributions, controlled by the physics below and above the
cut-off scale, respectively. Answer (I) would then require a
subtle (malicious?) conspiracy between low-energy and high-energy
physics, ensuring the desired fine-tuning. The criticism of (II)
has to do instead with the $d>4$ operators: in order to sufficiently
suppress the coefficients of the dangerous operators associated with
proton decay, FCNC, etc., the new physics at the cut-off scale
$\Lambda$ must have quite non-trivial properties!

\begin{figure}[htb]
\vspace{-1cm}
\centerline{
\epsfig{figure=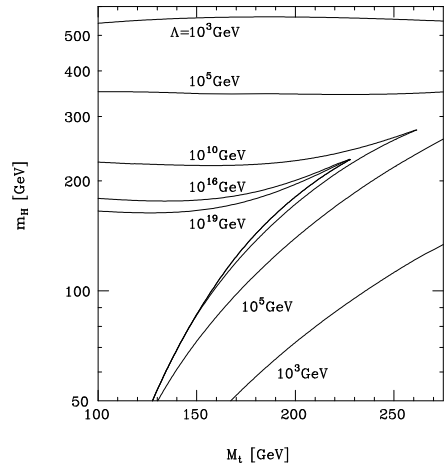,height=10cm,angle=90}}
\vspace{-1.0cm}
\caption{  Bounds in the $(M_t,m_H)$ plane, for various
           choices of $\Lambda$.}
\label{mariano}
\end{figure}

At the moment, answer (I) is not very popular in the physics
community, since we do not have the slightest idea on how the
required conspiracy could possibly work at the fundamental level.
Conceptually, such a possibility can be theoretically tested in
an ultraviolet-finite Theory of Everything: as daring as it may
sound, with the advent and the continuing development of string
theories, we may not be very far from the implementation of the
first quantitative tests. More concretely, such a possibility can
be experimentally tested in the near future, via the search for the
Higgs boson at LEP, at the Tevatron and at the LHC. A clear picture
of the implications of (I) is given in figure~\ref{mariano}, which
shows, for various possible choices of $\Lambda$ in the SM, the
values of the top quark and Higgs boson masses allowed by the
following two requirements\cite{cmpp}:
\begin{itemize}
\item
The SM effective potential should not develop, besides the minimum
corresponding to the experimental value of the electroweak scale,
other minima with lower energy and much larger value of the Higgs
field. In first approximation, this amounts to requiring the SM
effective Higgs self-coupling, $\lambda(Q)$, not to become negative
at any scale $Q < \Lambda$: for a given value of the top quark mass
$M_t$, this sets a lower bound on the SM Higgs mass $m_H$.
\item
The SM effective Higgs self-coupling should not develop a Landau
pole at scales smaller than $\Lambda$: for a given value of $M_t$,
this sets an upper bound on $m_H$. Such constraint has a meaning
which goes beyond perturbation theory, as suggested by the infrared
structure of the SM renormalization group equation for $\lambda(Q)$
and confirmed by explicit lattice computations\cite{lattice}.
\end{itemize}
Figure~\ref{mariano} includes some recent refinements\cite{higgswg}
of the original analysis, such as two-loop renormalization
group equations, optimal scale choice, finite corrections to the pole
top and Higgs masses, etc. For very large cut-off scales, $\Lambda
= 10^{16}$--$10^{19}$~GeV, the results are quite stable and can be
summarized as follows: for a top quark mass close to 180~GeV, as
measured at the Tevatron collider\cite{top}, the only allowed range
for the SM Higgs mass is 130~GeV $< m_H <$~200~GeV. This means that,
even in the absence of a direct discovery of new physics beyond the
SM, answer (I) could be falsified by LEP, the Tevatron and the LHC
in two possible ways: either by discovering a SM-like Higgs boson
lighter than 130~GeV, or by excluding a SM-like Higgs boson in the
130--200~GeV range!

Answer (II), instead, gives rise to a well-known conceptual
bifurcation:
\begin{description}
\item[{\bf (IIa)}]
In the description of electroweak symmetry breaking, the elementary
SM Higgs scalar is replaced by some fermion condensate, induced by
a new strong interaction near the Fermi scale. This includes old
and more recent variants of the so-called {\bf technicolor}
models\cite{techni} (`extended', `walking', `non-commuting', \ldots).
The stringent phenomenological constraints on technicolor models
coming from electroweak precision data will be mentioned later. On
the theoretical side, technicolor remains quite an appealing idea,
still waiting for a satisfactory and calculable model. The lack of
substantial theoretical progress in this field, however, may be due
to the technical difficulties of dealing with intrinsically
non-perturbative phenomena. This should not and certainly will not
prevent the experimentalists from keeping an open mind when looking
for possible signals of new physics.
\item[{\bf (IIb)}]
The SM is embedded in a model with softly broken global
supersymmetry,
and supersymmetry-breaking mass splittings between the SM particles
and their superpartners are of the order of the electroweak scale.
This approach, generically denoted as {\bf low-energy
supersymmetry}\cite{susy}, ensures the absence of field-dependent
quadratic divergences, and makes it `technically' natural that
there exists scalar masses much smaller than the cut-off scale.
Moreover, a minimal and calculable model is naturally singled out,
the so-called Minimal Supersymmetric Standard Model (MSSM).
\end{description}
\section{Extensions near the Fermi scale
(mainly MSSM phenomenology)}

This section reviews some phenomenological aspects of SM extensions
near the Fermi scale. Reflecting the content of the parallel sessions
and the personal taste of the speaker, most of it will deal with the
MSSM and its variants.

In order to set the framework for the following discussion, it is
useful to recall  the defining assumptions of the MSSM. The field
content is organized in gauge and matter multiplets of $N=1$
supersymmetry. The gauge group is $G=SU(3)_C \times SU(2)_L
\times U(1)_Y$, and the matter content corresponds to three
generations of quarks and leptons, as in the SM, plus two
complex Higgs doublets, one more than in the SM. To enforce
baryon- and lepton-number conservation in $d=4$ operators, one
imposes a discrete $R$-parity: $R=+1$ for all ordinary particles
(quarks, leptons, gauge and Higgs bosons), $R = -1$ for their
superpartners (spin-0 squarks and sleptons, spin-$1/2$ gauginos and
higgsinos). A globally supersymmetric Lagrangian ${\cal L}_{\rm
SUSY}$
is then fully determined by the superpotential (in standard
notation):
\be
\label{mssmw}
f = h^U Q U^c H_2 + h^D Q D^c H_1
+ h^E L E^c H_1 + \mu H_1 H_2 \, .
\ee

To proceed towards a realistic model, one has to introduce
supersymmetry breaking. In the MSSM, supersymmetry breaking
is {\em parametrized} by a collection of {\em soft} terms,
${\cal L}_{soft}$, which preserve the good ultraviolet
properties of global supersymmetry. ${\cal L }_{soft}$ contains
mass terms for scalar fields and gauginos, as well as a restricted
set of scalar interaction terms
\bea
\label{lsoft}
- {\cal L }_{soft}
& = &
\sum_i \tilde{m}_i^2 | \varphi_i|^2 + {1 \over 2}
\sum_A M_A \ov{\lambda}_A \lambda_A
\nonumber \\
& & \nonumber \\
& + & \left( h^U A^U Q U^c H_2 + h^D A^D Q D^c H_1
\right.
\nonumber \\
& & \nonumber \\
& + &
\left.
h^E A^E L E^c H_1 + m_3^2 H_1 H_2 + {\rm h.c.}
\right),
\eea
where $\varphi_i$ ($i=H_1,H_2,Q,U^c,D^c,L,E^c$) denotes the generic
spin-0 field, and $\lambda_A$ ($A=1,2,3$) the generic gaugino field.
Observe that, since $A^U,A^D$ and $A^E$ are matrices in generation
space, ${\cal L }_{soft}$ contains in principle a huge number of free
parameters. Moreover, for generic values of these parameters one
encounters phenomenological problems with FCNC, CP violation,
charge- and colour-breaking vacua. All the above problems can be
solved at once if one assumes that the running mass parameters in
${\cal L}_{soft}$, defined at the one-loop level and in a
mass-independent renormalization scheme, can be parametrized, at
a cut-off scale $\Lambda$ close to $\mpla$, by a universal gaugino
mass $m_{1/2}$, a universal scalar mass $m_0$, and a universal
trilinear scalar coupling $A$, whereas $m_3^2 \equiv - B \mu$
remains in general an independent parameter.

\subsection{MSSM (and alternatives) vs. electroweak precision
data}

The theoretical interpretation of electroweak precision data, in
the framework of the SM and of its candidate extensions (including
the MSSM), has been the subject of several talks in the
parallel\cite{ewprec} and plenary\cite{hollik,olchevski} sessions.

Universal effects, occurring via the vector-boson self-energies,
and parametrized in terms of convenient variables\cite{pt} such as
$(S,T,U)$ or $(\epsilon_1,\epsilon_2,\epsilon_3)$, have already been
discussed many times at this and previous conferences, and the
results can be summarized as follows:
\begin{itemize}
\item
The SM fits excellently all the data (with the value of the strong
coupling constant extracted from the hadronic $Z$ and $\tau$
branching ratios slightly higher than, but still compatible with,
the one extracted from deep-inelastic scattering).
\item
The MSSM gives at least as good a fit as the SM, thanks to the fast
decoupling properties of the virtual effects of supersymmetric
particles, as long as their mass is increased above the $m_Z/2$
threshold.
\item
Naive versions of technicolor and extended technicolor models are
ruled out (whereas some `walking' technicolor models may still work).
\end{itemize}

A point that has attracted increasing attention in the months before
this Conference is the fact that, in some extensions of the SM,
non-universal effects on the $Z b \overline{b}$ vertex are also
possible, which can modify appreciably the SM prediction for
$R_b \equiv \Gamma(Z \rightarrow b \overline{b})/\Gamma(Z
\rightarrow hadrons)$. In the MSSM, one can have\cite{bf} extra
positive contributions to $R_b$ from loops involving stop squarks
and charginos or bottom quarks and neutral Higgs bosons, extra
negative contributions to $R_b$ from loops involving the top quark
and the charged Higgs boson. In the technicolor framework, one
has\cite{cst} extra negative contributions to $R_b$ in `walking'
technicolor models, whereas contributions can be of either sign
in `non-commuting' technicolor models.

\begin{figure}[htb]
\vspace{-.1cm}
\centerline{
\epsfig{figure=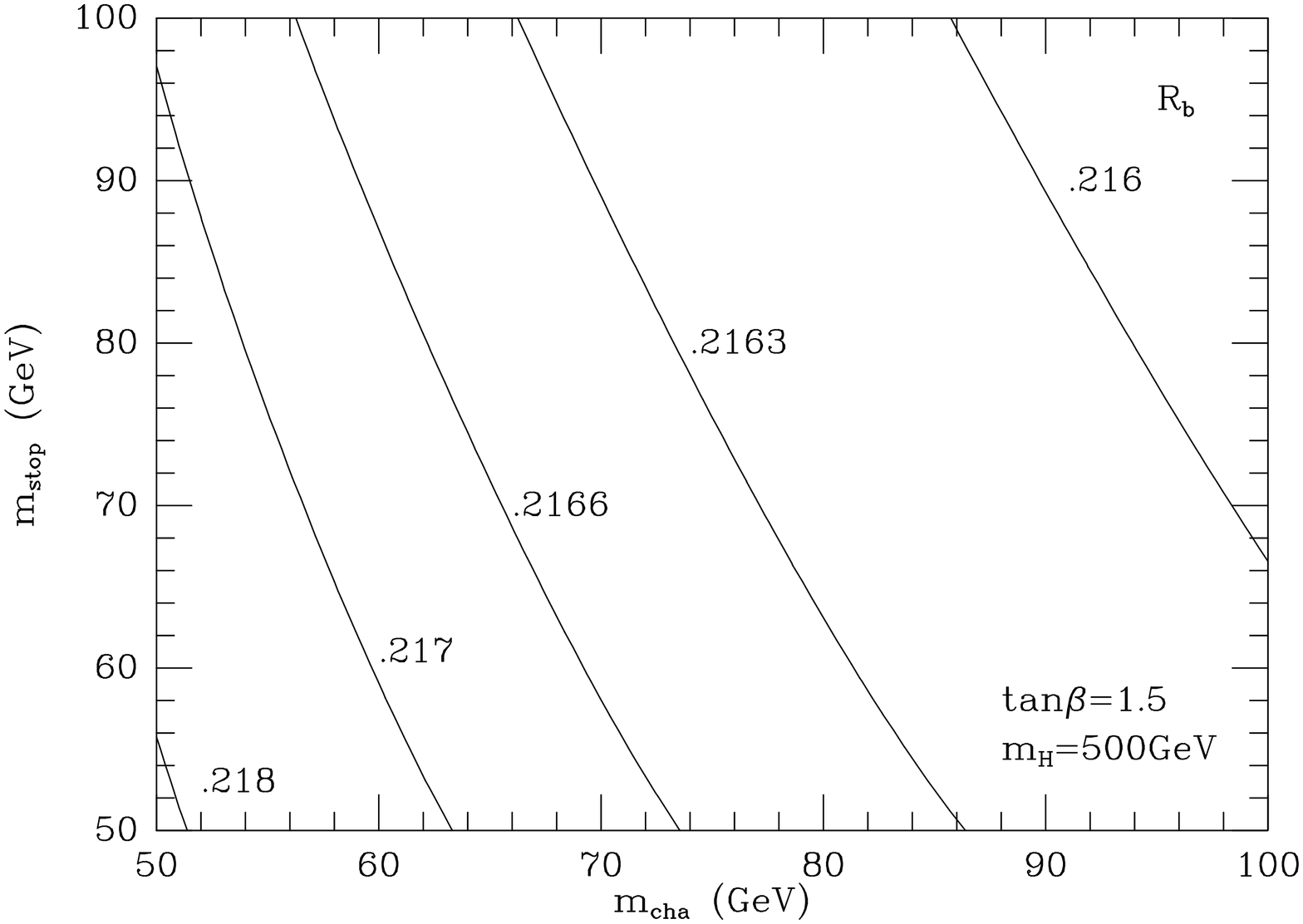,height=4.5cm,angle=-90}}
\vspace{2cm}
\caption{Contours of $R_b$ in the plane defined by the mass $m_{\rm
cha}$ of the lightest chargino (taken here to be $\tilde{H}^{\pm}$)
and the mass $m_{\rm stop}$ of the lightest stop squark (taken here
to be $\tilde{t}_R$).}
\label{bfzrb}
\end{figure}

The experimental data available before this Conference\cite{lep01}
suggested\cite{sugg} that an improved fit to $\alpha_S$ and
$R_b$ could be obtained in the MSSM in the case of light
stops and charginos (with generic $\tb$) and/or light
$A^0$ (with $\tb \sim m_t/m_b$). For the effect to be
numerically significant, the non-standard particles in the loops
should not be much heavier than $m_Z/2$, otherwise fast decoupling
would take place and the effect rapidly vanish. A quantitative
estimate of this effect is given\cite{bfzpheno} in
figure~\ref{bfzrb},
which includes, besides the standard $(t,W^{\pm})$ loop, also the
$(t,H^{\pm})$ loop and the $(\tilde{t},\tilde{\chi}^{\pm})$ loops,
in the simplified case of light $\tilde{t}_R$ and $\tilde{H}^{\pm}$.

After the new data presented at this Conference\cite{olchevski}, the
picture appears more confused. Now both $R_b$ and the analogous ratio
$R_c$ have been measured to better accuracy and with different
methods.
The theoretical SM prediction [for $M_t=180\pm12$~GeV,
$m_H=65$--$1000$~GeV, $\al_S(m_Z)=0.125\pm0.007$ and
$\al^{-1}(m_Z)=128.90\pm0.09$] and the averaged
experimental determinations are:
\vspace*{0.2cm}
\begin{center}
\begin{tabular}{|c|c|c|}
\hline
& & \\
& EXP. & TH.(SM) \\
& & \\
\hline
& & \\
$R_b$ & $0.2219 \pm 0.0017$ & $0.2156 \pm 0.0005$ \\
& & \\
\hline
& & \\
$R_c$ & $0.1540 \pm 0.0074$ & $0.1724 \pm 0.0003$ \\
& & \\
\hline
\end{tabular}
\end{center}
\vspace*{0.2cm}
We then have an excess in $R_b$ at about the $3.5\sigma$ level and
a defect in $R_c$ at about the $2.5\sigma$ level. Taking into account
the measured value of the total hadronic width,
\be
\Gamma_{h,exp} = (1744.8\pm3.0) \; {\rm MeV} \, ,
\ee
which comfortably agrees with the SM prediction,
\be
\Gamma_{h,SM} = (1745.7 \pm 6.0) \; {\rm MeV} \, ,
\ee
one finds the following discrepancies: $\delta \Gamma_b = 11
\pm 3$~MeV, $\delta \Gamma_c = -32 \pm 13$~MeV, $\delta (\Gamma_b
+ \Gamma_c) = -21 \pm 12$~MeV. Notice that the discrepancy in
$\Gamma_b+\Gamma_c$ is much larger than the error on $\Gamma_h$,
and in sign and magnitude cannot support any longer any intriguing
connection between the experimental effects on $\al_S$ and $R_b$.
Also, fitting the data within the MSSM now becomes impossible: for
stops, charginos and $A^0$ all around 50~GeV, and $\tb \sim m_t/m_b$,
the MSSM could marginally reproduce the observed value of $R_b$,
but the improvement in the fit to $R_c$ with respect to the SM
would be negligible.

One could then fix (somewhat arbitrarily) $R_c$ to its SM value.
In this case, the fit to the experimental data would give $R_b=0.2205
\pm 0.0016$, roughly $3 \sigma$ in excess of the SM prediction.
Still, as can be appreciated from figure~\ref{bfzrb}, the discrepancy
would be large enough that, barring very special regions of the
parameter space, which may be already ruled out by indirect
constraints
or soon ruled out by the forthcoming LEP run at
$\sqrt{s}=130$-$140$~GeV,
the MSSM can provide only a modest improvement in the quality of the
fit.

\subsection{MSSM and the decay $b \rightarrow s \gamma$}

As discussed in the parallel sessions\cite{ricciardi,carena}, the
recent experimental observation of radiative $B$ decays\cite{cleo}
plays today a very important role in constraining many extensions
of the SM, and in particular the MSSM.

\begin{figure}[htb]
\vspace{-.1cm}
\centerline{
\epsfig{figure=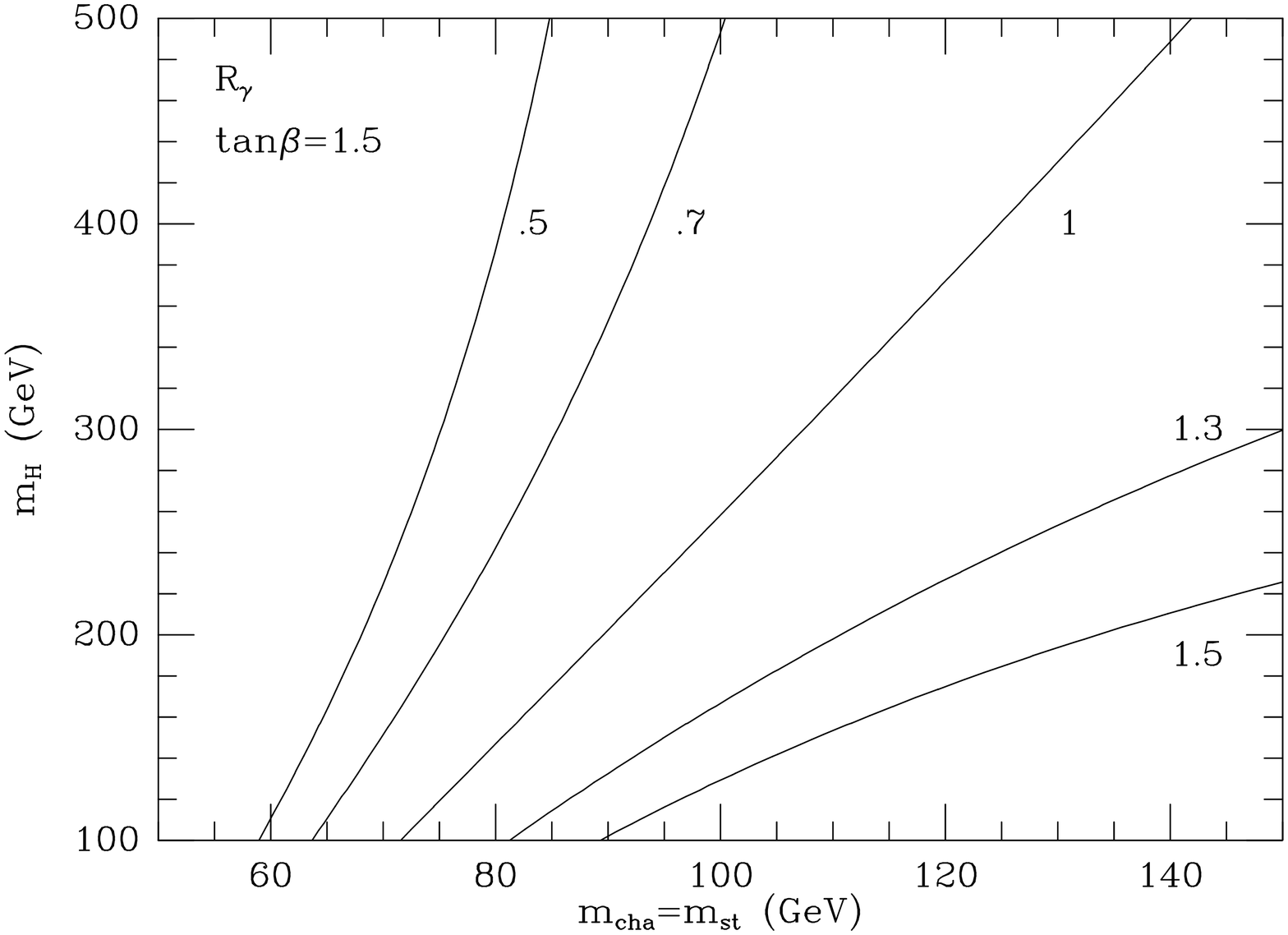,height=4.5cm,angle=-90}}
\vspace{2cm}
\caption{Contours of $R_{\gamma}$ in the $(m_{\rm cha}=
m_{\rm st},m_H)$ plane.}
\label{bfzrg}
\end{figure}

The experimental number most easily compared with theory is the
inclusive branching ratio
\be
\label{bsgexp}
\left[ BR ( B \rightarrow X_s \gamma ) \right]_{exp} =
( 2.32 \pm 0.67 ) \times 10^{-4} \, .
\ee
In the SM, this process is described, at the partonic level
$(b \rightarrow s \gamma)$ and at lowest order, by loop
diagrams with internal top and $W^{\pm}$ lines. However,
the theoretical determination of the inclusive branching
ratio suffers from large uncertainties, mainly due to the
QCD corrections, which at the moment have been calculated
only at leading order\cite{bsgqcd}. A conservative
estimate\cite{ali} gives a total theoretical error of
roughly $50 \%$:
\be
\label{bsgsm}
\left[ BR ( B \rightarrow X_s \gamma ) \right]_{th}^{SM} =
( 2.55 \pm 1.28 ) \times 10^{-4} \, ,
\ee
whilst other less conservative estimates give theoretical errors
as low as $30 \%$. The excellent agreement between the two
determinations (\ref{bsgexp}) and (\ref{bsgsm}) can be taken as
another piece of evidence for SM radiative corrections.
This is not yet at the level of a precision test, but
already represents an important constraint on possible
new physics at the electroweak scale. For example, in the MSSM
there are additional diagrams, corresponding to $(t,H^{\pm})$ and
$(\tilde{t},\tilde{\chi}^{\pm})$ exchange, which can give quite large
contributions to the rate\cite{bsgmssm}. For heavy supersymmetric
particles, the data disfavour a light charged Higgs. More generally,
a correlation is enforced between a light charged Higgs and light
stops and charginos, since one needs the right amount of negative
interference to fit the data. The situation is illustrated in
figure~\ref{bfzrg}, which displays\cite{bfzpheno} contour lines of
$R_{\gamma} \equiv BR ( B \rightarrow X_s \gamma )_{MSSM} / BR ( B
\rightarrow X_s \gamma )_{SM}$, in the plane characterized by
a common mass for the lightest stop and charginos (taken here to
be $\tilde{t}_R$ and $\tilde{H}^{\pm}$) and by the charged Higgs
mass, for the representative value $\tb = 1.5$.
The calculation of the next-to-leading QCD corrections, announced
in a parallel session\cite{ricciardi},  would allow a significant
reduction of the theoretical error, and thus greatly enhance the
constraints on the MSSM and on generic two-Higgs-doublet models.

\subsection{`Relaxed' MSSM}

Some of the assumptions defining the MSSM are plausible but not
really compulsory, even if they may find a justification in some
theoretical constructions going beyond the MSSM. When discussing
the phenomenology of low-energy supersymmetry, it is important to
keep an open mind and to study what happens when some of these
assumptions are relaxed.

Two possibilities were discussed in the parallel sessions. The
first one\cite{brokenr} consists in writing down the most general
renormalizable superpotential compatible with supersymmetry and
the SM gauge symmetry, which contains, besides the familiar MSSM
terms of eq.~(\ref{mssmw}), the additional terms
\be
\label{rbreakw}
\Delta w = \lambda Q D^c L +
\lambda' L E^c L +
\lambda'' U^c D^c D^c \, ,
\ee
where $\lambda$, $\lambda'$ and $\lambda''$ have to be interpreted
as three-index tensors in generation space. Novel analyses of the
phenomenological constraints on the $R$-parity violating couplings
of eq.~(\ref{rbreakw}) were discussed in the parallel
sessions\cite{sher,bhat}. New bounds from non-leptonic
B-decays\cite{sher}, Z physics at LEP\cite{bhat} and
D-decays\cite{bhat} were presented. It was also observed\cite{sher}
that, if enough third-generation fields are involved, some
baryon- and lepton-number violating terms in (\ref{rbreakw})
could coexist with couplings of order $10^{-1}$--$10^{-2}$.

The second possibility consists in allowing non-universal
soft supersymmetry-breaking terms. This hypothesis is
subject to very stringent constraints from FCNC, as
discussed in the parallel sessions\cite{savoy}. An
example is the decay $\mu \rightarrow e \gamma$, subject to the
strong experimental bound $BR(\mu \rightarrow e \gamma) <
5 \times 10^{-11}$. Off-diagonal slepton mass terms in generation
space, denoted here with the generic symbol $\delta m^2$, would
contribute to the above decay at the one-loop level, and the
previous limit roughly translates into $\delta m^2 / m_{\tilde{l}}^2
< 10^{-3}$--$10^{-5}$, if one assumes gaugino masses of the order
of the average slepton mass $m_{\tilde{l}}$ (a quite complicated
parametrization is needed to formulate the bound more precisely).
Similar constraints can be obtained by looking at the $\ko$--$\kb$,
$\bo$--$\bb$ systems and at other flavour-changing phenomena. It is
important to recall that all these bounds are naturally respected by
the strict MSSM, where the only non-universality in the squark and
slepton mass terms is the one induced by the renormalization group
evolution from the cut-off scale $\Lambda$ to the electroweak scale.
However, the same bounds represent quite non-trivial requirements on
extensions of the MSSM, such as supersymmetric grand-unified theories
(SUSY GUTs) and string effective supergravities, since in general one
expects non-universal contributions to the soft
supersymmetry-breaking
masses. Various mechanisms that could enforce the desired amount of
universality, or a sufficient suppression of FCNC via approximate
alignments of the fermion and sfermion mass matrices, have been
presented
in the mini-review by Savoy\cite{savoy}. Another interesting recent
development is the attempt\cite{barbieri} to establish a link,
in the framework of SUSY GUTs, between the magnitude of the top
quark mass and the amount of FCNC expected in the resulting,
`relaxed' version of the MSSM. In order to do so, one defines a
SUSY GUT, with universal soft mass terms, near the scale $\mpla$,
and follows the logarithmic renormalization group evolution of
the model parameters from $\mpla$ to $M_U$: the large top Yukawa
coupling controls the amount of non-universality generated at $M_U$.
A possible limit to the predictivity of this analysis is, in my
opinion, the assumption that the logarithmic RG evolution in the
$(M_U,\mpla)$ interval, with $\beta$-functions as computed in the
specific SUSY GUT model, is a good approximation. However, this
criticism does not spoil the interest of such an analysis: one
can introduce a general parametrization for the universality
violations at $M_U$, or, equivalently, at the electroweak scale,
and study the bounds on these parameters coming from FCNC processes;
these will have to be respected by any fundamental theory that
claims to predict the soft mass parameters of the MSSM.

\subsection{How could the MSSM be falsified?}

A legitimate question, often asked when searches for new
particles\cite{grivaz} are described, is the following:
How could the MSSM be falsified, in the absence of new
experimental discoveries?

Apart from the generic `naturalness' argument, requiring the
masses of supersymmetric particles to be of the order of the
Fermi scale, namely smaller than a few TeV, it is difficult
to establish firmer theoretical upper bounds. Attempts to quantify
an acceptable `measure of fine-tuning' and use it to bound from
above the supersymmetric particle masses\cite{ftun} are
parametrization-dependent, and should be taken just as
indications, since they do not have a solid theoretical
foundation.

\begin{figure}[htb]
\vspace{-0.1cm}
\centerline{
\epsfig{figure=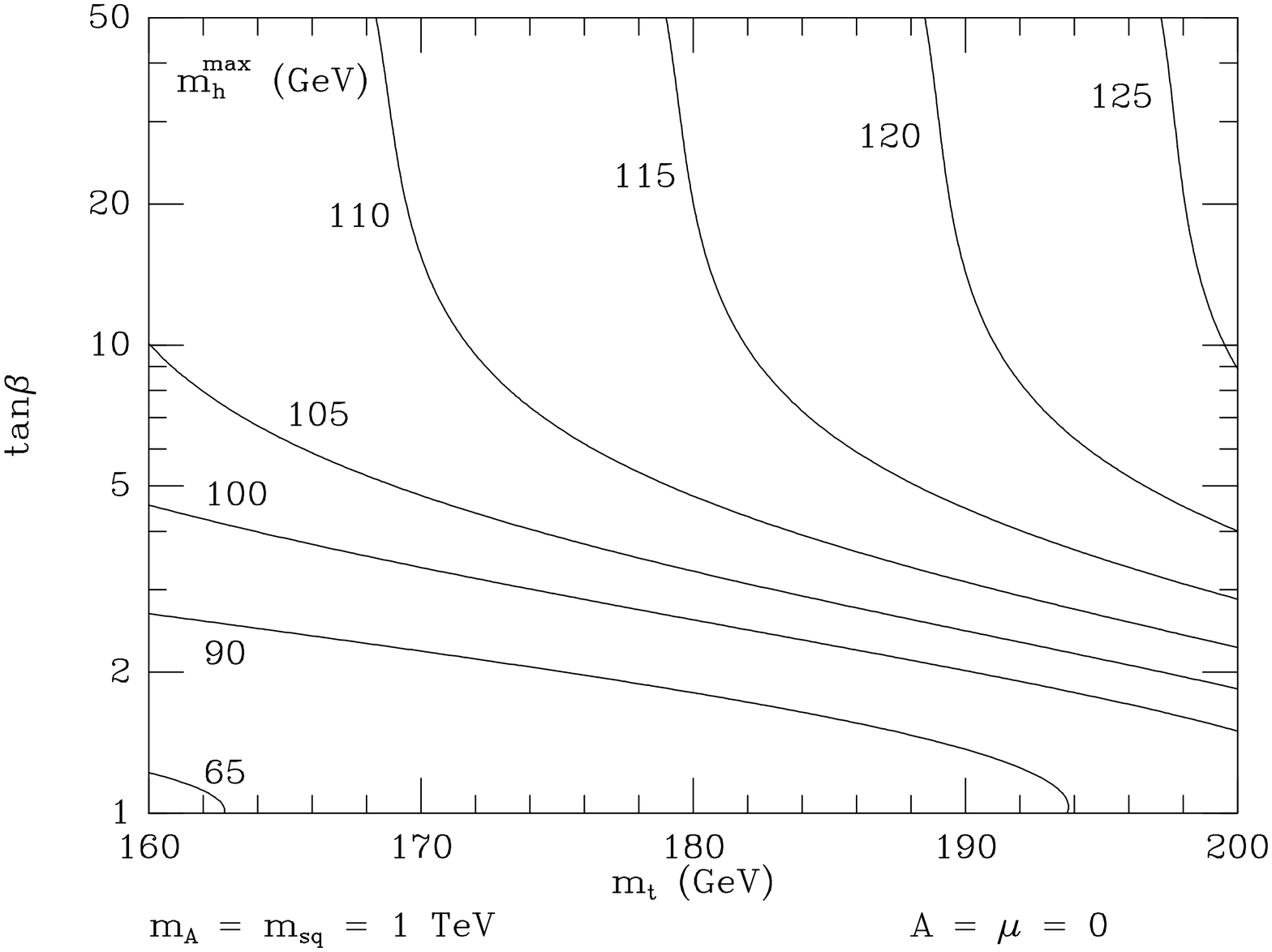,height=4.5cm,angle=-90}}
\vspace{2.0cm}
\centerline{
\epsfig{figure=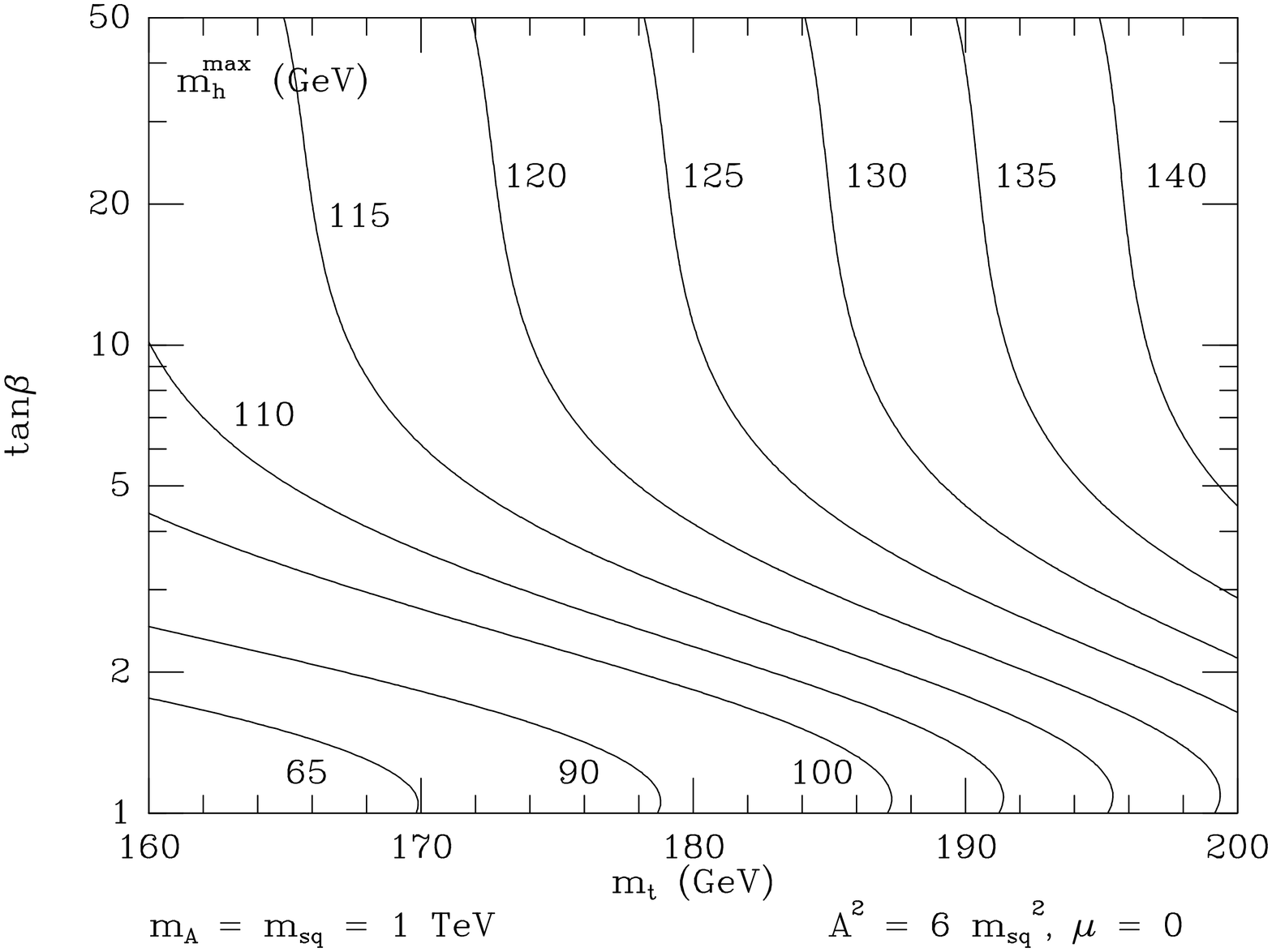,height=4.5cm,angle=-90}}
\vspace{2.0cm}
\caption{Contours of $m_h^{max}$  in the $(m_t,\tan
\beta)$ plane, for $m_{sq} = 1$ TeV and negligible
or maximal mixing.}
\label{kz1}
\end{figure}

However, the Higgs sector of the MSSM is very tightly constrained.
At the classical level, the mass of the lightest CP-even neutral
Higgs boson obeys the celebrated inequality $m_h < m_Z |\cos 2 \beta
|$. This bound is shifted by the radiative corrections\cite{erz}.
For example, the leading one-loop correction, due to the exchange of
the top quark and of its scalar partners, involves a shift in the
`22' diagonal entry of the CP-even mass matrix,
\be
(\Delta m^2)_{22} = {3 g^2 \over 8 \pi^2}
{m_t^4 \over m_W^2 \sin^2 \beta} \log {m_{\tilde{t}_1}
m_{\tilde{t}_2} \over m_t^2} + \ldots \, ,
\ee
which clearly exhibits the relevant dependences on the top and stop
masses. Further refinements in the calculation of the radiative
corrections to the MSSM Higgs masses, and in particular of the upper
bound $m_h^{\rm max}$ on $m_h$, include the parametrization of mixing
effects in the stop sector (which can give in some cases an extra
positive shift in $m_h$), the resummation of the leading logarithms
via the renormalization group (which in general decreases the upper
bound on $m_h$), the momentum-dependence of the self-energies and
loops of other MSSM particles (which give in general small effects).
The results of a state-of-the-art calculation\cite{higgswg} are
illustrated in figure~\ref{kz1}, which displays contours of
$m_h^{\rm max}$ in the $(m_t,\tb)$ plane, for large average stop mass
($m_{\rm sq} \equiv \sqrt{(m_{\tilde{t}_1}^2 +
m_{\tilde{t}_2}^2)/2}$)
and negligible or maximal mixing effects, respectively. It should
be stressed that $m_h^{\rm max}$ is the maximum possible value of
$m_h$, essentially saturated for $m_A = 1$~TeV, but not necessarily
the theoretically most probable value, since it is obtained by
pushing the MSSM parameters to the limits of their plausible range
of variation.

Similarly, only slightly weaker bounds can be established within
supersymmetric models with non-minimal Higgs sectors\cite{higgswg}.
Therefore, excluding the predicted Higgs sectors stands out as the
most promising option for falsifying the MSSM and its non-minimal
variants at future accelerators\cite{maiani}. Positive evidence
for supersymmetry, however, can only come from the discovery of
some (R-odd) supersymmetric particle.

\subsection{`Constrained' MSSM}

A remarkable fact, extensively advertised in the last few years,
is the following: combining the extracted values of the effective
gauge couplings at the weak scale and the leading logarithmic
evolution of the latter\cite{gqw} in the MSSM (with no new
thresholds), one gets a consistent picture of approximate
unification of the gauge couplings at a scale $M_U \sim 2
\times 10^{16}$~GeV.

This stunning success, however, does not allow us to single out
a unique SUSY GUT replacing the MSSM at the scale $M_U$! In
constructing such a theory, there is freedom to choose the unified
gauge group, the representations in the Higgs sector, the parameters
of the superpotential couplings (including, in general, a number
of explicit mass terms), the structure of the soft terms after
spontaneous supersymmetry breaking. Even choosing the simplest
and most famous SUSY GUT, minimal SUSY SU(5)\cite{dg},
predictivity is limited by the freedom to choose the masses of
some of the heavy Higgs multiplets, and by the likely existence
of corrections to the SUSY-GUT Lagrangian, in the form of
little-suppressed non-renormalizable operators, induced by
physics at possible nearby scales (compactification scale,
string scale, Planck scale). Moreover, minimal SUSY SU(5)
must certainly be modified to incorporate a realistic fermion
mass spectrum and to solve the doublet--triplet splitting problem.

The moral of the story is that, when performing phenomenological 
analyses, it may be dangerous to put bounds on the MSSM mass 
spectrum by imposing additional constraints such
as `strict' gauge coupling unification, `strict'
bottom--tau Yukawa coupling  unification, proton decay
as described by minimal SUSY SU(5), or radiative electroweak
symmetry breaking with universal soft scalar masses at $M_U$.
Many strong (and indeed unnecessary) model dependences are
introduced! Before going to this level of detail, one would
need a believable theory at the scale $M_U$ and, in my opinion,
we have not yet reached such a stage. Therefore, some of the 
interesting analyses presented in the parallel
sessions\cite{carena,arnowitt,deboer}, technically correct
within their assumptions, must be interpreted with a grain 
of salt!

\section{Extensions near the Planck scale
(superstrings and their possible low-energy implications)}

In the search for a more fundamental theory going beyond
the MSSM, and allowing us to predict some of its many
parameters, we have today a great advantage with respect to
the early eighties, since we can make use of the impressive
progress of string theories over the last decade.

Superstrings\cite{strings} (perhaps to be replaced, some
day, by the conjectured `M-theory', of which the various
string theories may be different perturbative expansions)
are the only known candidate for a consistent,
ultraviolet-finite quantum theory of gravity, unifying all
fundamental interactions. There are perturbatively stable
four-dimensional solutions of the heterotic string with
nice phenomenological properties such as $N=1$ supersymmetry
in flat four-dimensional space-time, a gauge group $G$ containing
the SM gauge group $SU(3) \times SU(2) \times U(1)$, three
chiral families (and possibly extra stuff), and more.
Incidentally, the fact that supersymmetry seems to play a very
important role for the quantum stability of superstring vacua
may be taken as an additional motivation to favour low-energy
supersymmetry over technicolor: however, it should be kept in
mind that so far superstrings have not been able to give us any
definite insight about the scale of supersymmetry breaking.

The general feature to be stressed is that string theories contain
one explicit mass scale, the string scale, which fixes a mass unit
and acts as a physical ultraviolet cut-off. All the other physical
scales ($\mpla,M_U,m_Z,\ldots,$ in realistic models), and all the
dimensionless couplings of the low-energy effective theory (probably
some version of the MSSM), are controlled by the VEVs of some scalar
fields, called {\bf moduli}, corresponding, in the effective
supergravity theories, to perturbatively flat directions of the
scalar potential. The  inclusion of non-perturbative quantum effects
is expected to spontaneously break supersymmetry and to remove the
degeneracy in the moduli space, thus selecting the correct vacuum.

The special duality properties of string theories\cite{duality}
(some of which have their counterpart at the field-theory level,
as discussed at this Conference by E.~Verlinde\cite{eric}) can
play a crucial role in controlling these phenomena. The 
best-known string dualities are the so-called $T$-dualities, of 
which the simplest example is the equivalence between a string
compactified on a circle of radius $R$ and the same string
compactified on a circle of radius $1/R$ (in appropriate string
units). These dualities are perturbative, in the sense that the
duality transformations do not act on the dilaton field, whose VEV
controls the coupling constant associated with the string loop
expansion, so they can be consistently defined in the weak-coupling
limit. The dualities at the origin of a lot of recent excitement
are however the so-called $S$-dualities, which interchange weak
and strong coupling, and are therefore inherently non-perturbative.
As explained by Verlinde, a prototype of $S$-duality is the
well-known electric--magnetic duality of QED. In supersymmetric
theories, electric--magnetic duality is expected to be part of a
larger set of transformations, acting both on the gauge coupling
$g$, controlling the $F^2$ term in the Lagrangian, and on the
vacuum angle $\theta$, controlling the associated $F \tilde{F}$
term, combined into a single chiral superfield $S$. There is mounting
evidence that $S$-duality is indeed a symmetry of the ten-dimensional
heterotic string compactified on a six-torus, as well as of globally
supersymmetric $N=4$ Yang-Mills theories. Even more interestingly,
examples are being found of dual pairs of string theories, in which
one string theory at strong coupling is equivalent to another string
theory at weak coupling. Most of the evidence collected so far
concerns
string theories that would have unbroken $N>1$ supersymmetry in
$d=4$,
but the physically most important goal is clearly to understand the
theories with $N=1$ and $N=0$ supersymmetries in four dimensions: it
would be great if one could study non-perturbative phenomena in
realistic string models just by going to the dual, weakly-coupled
theory!  Important conceptual developments are rapidly taking place
also in this respect. Waiting for solid results, applicable to
realistic cases, we are already witnessing a change of perspective
in the approach to some phenomenological problems.
In the rest of this talk, I would like to mention some of them,
not because they are particularly important, but because they are
the ones in which I have recently been involved.

\subsection{Supersymmetry breaking}

At the level of dimensionless couplings, the MSSM is more
predictive than the SM, since its quartic scalar couplings
are related by supersymmetry to the gauge and the Yukawa
couplings. The large amount of arbitrariness
in the MSSM phenomenology is strictly related to its
explicit mass parameters, the soft supersymmetry-breaking
masses and the superpotential Higgs mass. Such arbitrariness
cannot be removed within theories with softly broken global
supersymmetry, such as SUSY GUTs: to make progress,
{\em spontaneous} supersymmetry breaking must be introduced.

To discuss spontaneous supersymmetry breaking in a realistic 
and consistent framework, gravitational interactions cannot 
be neglected. One is then led to $N=1$, $d=4$ supergravity, 
seen as an effective theory below the Planck scale, within 
which tree-level calculations can be performed and some 
qualitative features of the ultraviolet-divergent one-loop 
quantum corrections be studied. Of course, infrared 
renormalization effects can be studied, but they are plagued by 
the ambiguities due to the counterterms for the renormalizable
operators. To proceed further, one must go to $N=1$, $d=4$
superstrings, seen as realizations of a fundamental
ultraviolet-finite theory, within which quantum corrections
to the low-energy effective action can be consistently taken
into account, with no ambiguities due to the presence of
arbitrary counterterms.

In recent years, two approaches to the problem have been
followed. On the one hand, four-dimensional tree-level
string solutions, in which $N=1$ local supersymmetry is
spontaneously broken via orbifold compactifications, have
been constructed\cite{ss}: none of the existing examples is
fully realistic, however they represent a useful laboratory
to perform explicit and unambiguous string calculations.
On the other hand, many studies have been performed
within string effective supergravity theories, assuming
that supersymmetry breaking is induced by non-perturbative
phenomena such as gaugino condensation\cite{gcond}:
the loss in predictivity is compensated by the possibility
of a more general parametrization, including
non-perturbative effects that are still hard to handle
at the string theory level. With the advent of string--string
dualities, it is even conceivable that the two approaches may
be related (in an interesting paper that appeared after this
Conference\cite{ks}, it is argued that string tree-level
breaking in a type~II string solution may be dual to
non-perturbative breaking in a heterotic counterpart).

Before proceeding with the discussion, it may be useful to
recall some basic facts of $N=1$, $d=4$ supergravity\cite{cremmer}.
The theory can be formulated with three types of supermultiplets:
in addition to the chiral and vector supermultiplets, already
present in global supersymmetry, we need to introduce the
gravitational supermultiplet, whose physical degrees of freedom
are the spin-2 graviton and its supersymmetric partner, the
spin-$3/2$ gravitino. Up to higher-derivative terms, the theory
is completely determined by two functions of the chiral superfields:
one is the K\"ahler function ${\cal G} (z, \overline{z})
= K (z, \overline{z}) + \log |w(z)|^2$, which controls the kinetic
terms and the interactions of the chiral multiplets; this function
is conventionally decomposed into a K\"ahler potential $K$ and a
superpotential $w$. The other is the gauge kinetic function
$f_{ab}(z)$,
which controls the kinetic terms and the interactions of the vector
supermultiplets. It is customary to work in the natural supergravity
units, where all masses are expressed in units of the Planck mass,
i.e. $\mpla=1$ by convention. An important difference with global
supersymmetry is that the scalar potential is no longer
positive-semidefinite, but takes the form
\be
\label{sugpot}
V_0 = |D_a|^2 + |F_i|^2 - 3 e^{\cal G} \, ,
\ee
where the first two terms are positive-semidefinite, in analogy
with the usual F- and D-term contributions of global supersymmetry,
whereas the last term, associated with the auxiliary field of the
gravitational supermultiplet, is negative-definite. The novel
structure of the potential in supergravity theories permits the
breaking of  supersymmetry with vanishing vacuum energy, if the last
term in eq.~(\ref{sugpot}) cancels exactly the remaining ones
at the minimum: the order parameter for the breaking of local
supersymmetry in flat space is the gravitino mass, $m_{3/2}^2
= e^{\cal G}$, which fixes the scale of all supersymmetry-breaking
mass splittings, and therefore of the MSSM soft mass terms in the
low-energy limit.

The generic problems to be solved by a satisfactory
mechanism for spontaneous supersymmetry breaking
can be succinctly summarized as follows:
\begin{itemize}
\item
{\bf Classical vacuum energy.}
The potential of $N=1$ supergravity does not have a definite
sign and scales as  $m_{3/2}^2 \mpla^2$: already at the classical
level, one must arrange for the vacuum energy to be vanishingly
small with respect to its natural scale.
\item
{\bf ($m_{3/2}/\mpla$) hierarchy.}
In a theory where the only explicit mass scale is the reference
scale $\mpla$ (or the string scale), one must find a convincing
explanation of why it is $m_{3/2} \simlt 10^{-15} \mpla$ (as
required by a natural solution to the hierarchy problem), and
not $m_{3/2} \sim \mpla$.
\item
{\bf Stability of the classical vacuum.}
Even assuming that a classical vacuum with the above properties
can be arranged, the leading quantum corrections to the effective
potential of $N=1$ supergravity scale again as $m_{3/2}^2 \, \mpla^2$,
too severe a destabilization of the classical vacuum to allow for
a predictive low-energy effective theory.
\item
{\bf Universality of squark/slepton mass terms.}
Such a condition (or alternative but equally stringent ones)
is phenomenologically necessary to adequately suppress
FCNC, but is not guaranteed in the presence of general
field-dependent kinetic terms.
\end{itemize}
{}From the above list, it should already be clear that the
generic properties of $N=1$ supergravity are not sufficient
for a satisfactory supersymmetry-breaking mechanism. Indeed,
no fully satisfactory mechanism exists, but interesting
possibilities arise within string effective supergravities.
The best results obtained so far are listed below:
 \begin{itemize}
\item
It is possible to formulate supergravity models where the classical
potential is manifestly positive-semidefinite, with a continuum of
minima corresponding to broken supersymmetry and vanishing vacuum
energy, and the gravitino mass sliding along a flat direction
\cite{noscale,noscqc}. A recent development is the construction
of models of this type where gauge and supersymmetry breaking 
are simultaneously realized, with goldstino components along 
gauge-non-singlet directions\cite{bzbfz}.
\item
This special class of supergravity models emerges naturally, as
a plausible low-energy approximation, from four-dimensional
string models, irrespectively of the specific dynamical mechanism
that triggers supersymmetry breaking. Due to the special geometrical
properties of string effective supergravities, the coefficient of
the one-loop quadratic divergences in the effective theory, $\str
\cm^2$, can be written as \cite{lhc}
\be
\label{genstr}
\str \cm^2 \zzbar = 2 \, Q \, m_{3/2}^2 \zzbar
\, ,
\ee
where $Q$ is a field-independent coefficient, calculable from the
modular weights of the different fields belonging to the effective
low-energy theory, i.e. the integer numbers specifying their
transformation properties under the relevant duality. The
non-trivial result is that the only field-dependence of
$\str \cm^2$ occurs via the gravitino mass. Since all
supersymmetry-breaking mass splittings, including those of the
massive string states not contained in the effective theory, are
proportional to the gravitino mass, this sets the stage for a natural
cancellation of the ${\cal O} (m_{3/2}^2 \, \mpla^2)$ one-loop
contributions to the vacuum energy. Indeed, there are explicit
string examples that exhibit this feature. If this property can
persist at higher loops (an assumption so far), then the hierarchy
$m_{3/2} \ll \mpla$ can be induced by the logarithmic corrections
due to light-particle loops\cite{noscqc}.
\item
In this special class of supergravity models one naturally obtains,
in the low-energy limit where only renormalizable interactions are
kept, very simple mass terms for the MSSM states $(m_0,m_{1/2},\mu,
A,B$ in the standard notation), calculable via simple algebraic
formulae from the modular weights of the corresponding fields
and easily reconcilable with the phenomenological universality
requirements\cite{lhc}. This last result can indeed be obtained
also in a slightly less restrictive framework\cite{soft}.
\end{itemize}

Just to give the flavour of the argument, we present here an
ultra-simplified example, which retains the relevant qualitative
features of the general case, without its full technical complexity.

Consider a supergravity theory containing as chiral superfields
a gauge-singlet $T$ (to be thought of as one of the superstring
moduli fields), and a number of charged fields $C^{\al}$ (to be
thought of as the matter fields of the MSSM and possibly others),
with K\"ahler potential
\be
K= - 3 \log ( T + \overline{T}) + \sum_{\al} |C^{\al}|^2
( T + \overline{T})^{\lambda_{\al}} + \ldots \, ,
\ee
and superpotential
\be
w_{SUSY} = d_{\al \beta \gamma} C^{\al} C^{\beta} C^{\gamma}
\, .
\ee
The model exhibits a classical invariance under the
following set of transformations, parametrizing the
continuous group $SL(2,R)$:
\be
\label{sl2r}
T \rightarrow {a T - i b \over i c T + d} \, ,
\;\;\;\;
C^{\al} \rightarrow (i c T + d)^{\lambda_{\al}} C^{\al} \, ,
\;\;\;\;
(ab-cd=1) \, .
\ee
The above symmetry can be interpreted as an approximate
low-energy remnant of a $T$-duality invariance under the
discrete group $SL(2,Z)$, corresponding to the restriction
of the transformations (\ref{sl2r}) to the case of integer
$(a,b,c,d)$ coefficients, and generated by the two
transformations $T \rightarrow 1/T$ and $T \rightarrow
T + i$. One can think of this $SL(2,Z)$ as an exact quantum
symmetry of the underlying string model. In the language
of supergravity, the K\"ahler potential transforms as
$K \rightarrow K + \phi + \overline{\phi}$, where $\phi$
is an analytic function, and the superpotential as
$ w \rightarrow w \exp ( - \phi)$, so that the full
K\"ahler function ${\cal G}$ remains invariant.

Without specifying the dynamics which induces the spontaneous
breaking of local supersymmetry, one can try to parametrize
the latter with a superpotential modification of the form
\be
w = w_{SUSY} + \Delta w \, ,
\;\;\;\;\;
\Delta w = k \ne 0 \, ,
\ee
where $k$ is a constant, independent of the modulus
field $T$, which can be thought of as the large-$T$
limit of a modular form of $SL(2,Z)$. In the case in which
other moduli fields are present, such as the dilaton--axion
field $S$ associated with the gauge coupling constant, one
can replace $k$ with a suitable function of $S$, with the
correct transformation properties under a possible $S$-duality.
Notice that the superpotential modification introduced above
breaks the invariance under $T \rightarrow 1/T$, but preserves
the shift symmetry $T \rightarrow T + i \alpha$. A low-energy
structure equivalent to the one introduced here has been found
in explicit constructions of string orbifold models with string
tree-level breaking\cite{ss}, but these results could have
more general validity, and apply also, with the appropriate
modifications, to the case of non-perturbative breaking.

In the supergravity theory defined above, by applying the
standard formalism one can easily verify the following
results:
\begin{itemize}
\item
Thanks to the identity $|F_T|^2 \equiv 3 e^{\cal G}$, the
scalar potential of eq.~(\ref{sugpot}) is automatically
positive-semidefinite. At any minimum of the potential
supersymmetry is broken and the gravitino mass, $m_{3/2}^2
= k^2/(T + \overline{T})^3 \ne 0$ if one takes for simplicity
$C^{\al}=0$, is classically undetermined. The modulus field
$T$ corresponds to a flat direction, as in the no-scale
models\cite{noscale}, and its fermionic partner $\tilde{T}$
plays the role of the goldstino in the super-Higgs mechanism.
\item
$\str \cm^2$ can be put in the form of eq.~(\ref{genstr}), with
\be
\label{charge}
Q = - 2 + \sum_{\al} ( 1 + \lambda_{\al}) \, ,
\ee
where the first addendum is the contribution of the massive
gravitino and the second one the contribution of the matter
fields.
\item
In MSSM notation, the following very simple mass terms are 
generated:
\be
{(m_0^2)_{\al} \over m_{3/2}^2} = 1 + \lambda_{\al} \, ,
\ee
\be
{(A)_{\al \beta \gamma} \over m_{3/2}} = 3 + \lambda_{\al}
+ \lambda_{\beta}+ \lambda_{\gamma} \, ,
\ee
\be
{(\mu)_{\al \beta} \over m_{3/2}} = 1 + {\lambda_{\al}
+ \lambda_{\beta} \over 2}  \, ,
\ee
\be
{(B)_{\al \beta} \over m_{3/2}} = 2 + {\lambda_{\al}
+ \lambda_{\beta} \over 2}  \, .
\ee
\end{itemize}
The above example can be easily generalized to include gauge
interactions, with a non-trivial moduli dependence of the
gauge kinetic function: non-vanishing gaugino masses can then
be generated, proportional to the gravitino mass, and
eq.~(\ref{charge}) can be modified accordingly. It is
important to stress that, in this framework, the phenomenologically
desirable universality properties of the soft mass terms can 
naturally arise as a consequence of $T$-duality. Furthermore, 
a non-vanishing $\mu$-term can be generated for the MSSM, 
proportional to $m_{3/2}$, even if the supergravity superpotential 
does not contain any explicit Higgs mass term.

The weakest point of the above construction is the absence of
a string calculation showing that, if there is cancellation of
the ${\cal O} ( m_{3/2}^2 \mpla^2)$ contributions to the
effective potential at one loop, this cancellation can persist
at higher loops. Since in the effective theory one can identify
some quadratically divergent two-loop graphs\cite{bpr}, such an
assumption is far from obvious. However, there are hints\cite{lhc}
that the numerical coefficient of eq.~(\ref{charge}) might be given
a topological interpretation, so such an assumption is not completely
arbitrary.

Under the assumption that no terms ${\cal O} ( m_{3/2}^2 \mpla^2)$
are generated by string quantum corrections to the effective 
potential, the possibility arises of treating the gravitino mass 
$m_{3/2}$ as a dynamical variable of the low-energy theory valid 
near the electroweak scale, namely the MSSM. Then the actual 
magnitude of the gravitino mass could be determined by the 
logarithmic quantum corrections\cite{noscqc}, as computed in the 
MSSM. The minimization condition of the one-loop effective potential 
$V_1$, with respect to $m_{3/2}$, would take the form\cite{kpz}:
\be
m_{3/2}^2 {\partial V_1 \over \partial m_{3/2}^2} = 2 V_1
+ {\str \cm^4 \over 64 \pi^2} = 0 \, .
\ee
The above equation can be interpreted as defining an infrared
fixed point for the vacuum energy, with the two terms in the
second member representing the canonical scaling and the
scaling violation by quantum corrections, respectively.
One can show that, for reasonable values of the boundary
conditions on the dimensionless parameters, an exponentially
suppressed hierarchy $m_{3/2} \ll \mpla$ can be generated.

Of course, the reason why $m_{3/2}$ can be treated as a
dynamical variable in the effective low-energy theory is
the existence of a very flat direction for the
modulus on which it depends monotonically. This means
that, after the inclusion of the ${\cal O} (m_{3/2}^4)$
quantum corrections, there will be some very light
gauge-singlet spin-0 fields, with `axion-like' or
`dilaton-like' couplings and masses ${\cal O} (m_{3/2}^2
/ \mpla)$, i.e. in the $10^{-3}$--$10^{-4}$~eV range if
$m_{3/2}^2 \sim G_F^{-1}$, with interesting astrophysical
and cosmological implications, including a number of
potential phenomenological problems\cite{sarkar}.

\subsection{Infrared moduli physics and the flavour problem}

Once the taboo has been broken, by considering a parameter of
the MSSM (in the previous example, the overall scale of its
mass terms) as a dynamical variable at the electroweak scale,
and some partial success obtained (a possible explanation for
the $m_{3/2} \ll \mpla$ hierarchy, at the price of one important
assumption and some unsolved cosmological problems), it is not
a big step to generalize the game to other MSSM parameters,
to see if there is a chance that other problems can be solved.

For example, in the case of non-universal soft mass terms one
has in general a very severe problem with FCNC\cite{savoy},
unless one can find a good reason to justify the alignment
of the quark and squark mass matrices. It has recently been
proposed\cite{dimgiu} that also the relative angles between
the quark and squark mass matrices may be considered as
dynamical variables: again, minimization of the vacuum energy
can induce at least partial alignment. Analogous considerations
have also been made\cite{ross} by considering a possible dynamics
just above the SUSY-GUT scale $M_U$.

Along similar lines, two groups\cite{kpz,binetruy,kprz}
have considered the possibility of treating some of the
Yukawa couplings of the MSSM as dynamical variables, as
reviewed in a parallel session\cite{dudas}. The goal is
to find a dynamical explanation for the numerical values
of some of the fermion masses, for example those of the
third-generation quarks.

\begin{figure}[htb]
\vspace{-3.cm}
\centerline{
\epsfig{figure=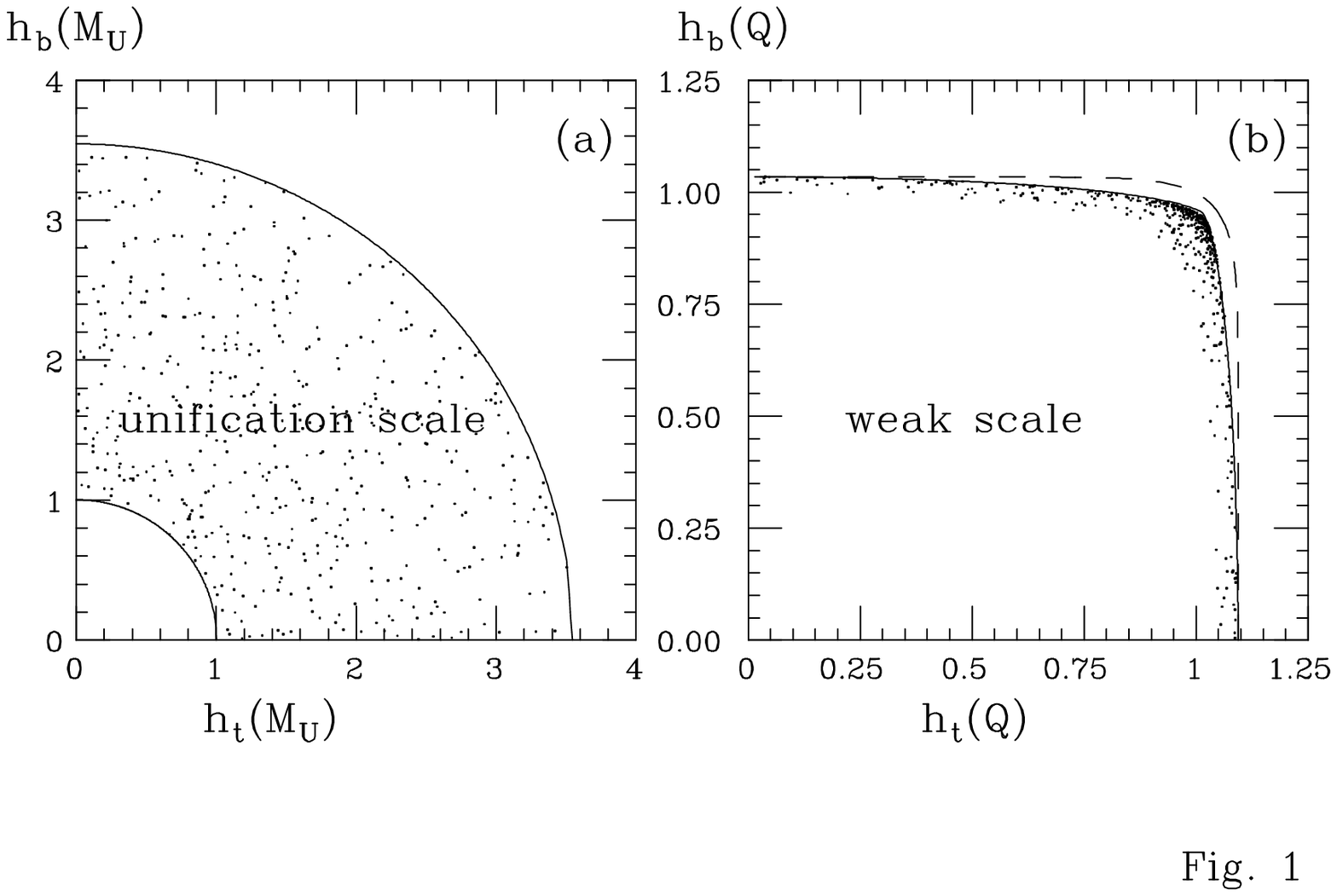,height=12cm,angle=0}}
\vspace{-3.0cm}
\caption{Mapping of the $(h_t(M_U),h_b(M_U))$ plane into the
$(h_t(Q),h_b(Q))$ plane, for $Q=200$~GeV. In (b), the dots correspond
to the exact numerical solutions of the one-loop RGE, for the
boundary
conditions given in (a); the solid and dashed lines correspond to two
different approximate analytical solutions.}
\label{yukkprz}
\end{figure}

To review the logic of the argument, we begin by recalling
that, in the MSSM, the RGEs for the top and bottom Yukawa
couplings admit an effective infrared fixed curve, analogous to
the effective infrared fixed point that one obtains\cite{ifp}
by setting the bottom Yukawa coupling to zero. Neglecting
for simplicity the $\tau$ Yukawa coupling and the electroweak
gauge couplings, an approximate analytical equation for the
infrared fixed curve is\cite{curve,kprz}
\be
\alpha_t + \alpha_b \simlt {2 \pi E \over 3 F} \cdot
f \left[ {4 \al_t \al_b \over (\al_t + \al_b)^2} \right]
\, ,
\ee
where to a good approximation $(2 \pi E / 3 F) \simeq (8/9)
\al_S$ and $f$ is a hypergeometric function bounded by
$1 \le f \le 12/7$. This infrared behaviour is
illustrated\cite{kprz} in figure~\ref{yukkprz}.
The previous considerations are only sufficient to set an
upper bound, of order 200~GeV, on the combination $m_t^2/
\sin^2 \beta + m_b^2 / \cos^2 \beta$, but cannot predict
the values of the top and bottom quark masses.

However, if after supersymmetry breaking some very flat
directions are left over in moduli space, and the Yukawa
couplings have some functional dependence on the corresponding
fields, then also the top and bottom Yukawa couplings can
be treated as dynamical variables of the low-energy theory.
A point stressed in a parallel session is that, in this
two-variable problem, the details of the moduli-dependence
of the two Yukawa couplings may or may not impose some
constraints on the minimization problem to be solved.
Irrespectively of these details, it can be shown that
minimization of the vacuum energy almost invariably brings
the low-energy couplings very close to the infrared fixed
curve of fig.~\ref{yukkprz}b. Unconstrained minimization,
however, favours the solution $h_t = h_t^{max} \gg h_b = 0$.
Constrained minimization, instead, can produce the phenomenologically
desired solution $h_t \simeq h_t^{max} \gg h_b \ne 0$, provided
that some constraint in the moduli space forbids the configuration
$h_b(M_U)=0$. Barring these model-dependent details, which could
be worked out only in a specific string construction, one is still
left with a definite prediction
\be
(M_t^{IR})^2 \simlt {m_t^2 \over \sin^2 \beta} +
{m_b^2 \over \cos^2 \beta} \simlt {12 \over 7}
(M_t^{IR})^2 \, ,
\ee
where $M_t^{IR} \simeq (4/3)\sqrt{\alpha_3/(\alpha_2+\alpha')}
m_Z \simeq 195$~GeV.
It seems difficult to go beyond the above result without
a more detailed knowledge of the moduli-dependence of Yukawa
couplings in string theories and of the mechanism for
spontaneous supersymmetry breaking.

\section{Conclusions}

Among the SM extensions at the Fermi scale, the MSSM stands out
as the theoretically most motivated and the phenomenologically
most viable one. In view of its limitations, as far as uniqueness
and predictivity are concerned, the MSSM should be taken as a
useful phenomenological parametrization, with the hope that its
many parameters will be fixed in the not-too-distant future by
the discovery of supersymmetric particles at LEPII, Tevatron and
the LHC. Paraphrasing an expression used in the parallel sessions,
this would be the real `bottom-up approach'!

On the more theoretical side, intense study over the last ten
years increasingly suggests that strings must be taken seriously
as a candidate fundamental theory defined near the Planck scale.
The connection between string theories and the MSSM, which we would
like to understand as the low-energy effective field theory near the
correct string vacuum, is made difficult by the problem of
spontaneous
supersymmetry breaking. There is great hope that we may soon extend
the exciting results on non-perturbative dualities in supersymmetric
field theories and string theories to more and more realistic
situations,
and promising lines of development have continued to flourish in the
months between the end of this Conference and the preparation of this
written contribution.

To conclude the talk, I would like to express my personal belief:
both
in experiment and in theory, we are heading to very exciting times!
\setcounter{secnumdepth}{0} 
%
\section{Acknowledgements}
In the preparation of section~2.1, I have benefited from discussions
with G.~Altarelli. I would also like to thank A.~Brignole and 
C.~Kounnas for some useful comments.
\end{document}